\documentclass[aps,pra,showpacs,amsmath,amssymb,amsfonts,lengthcheck,twocolumn,longbibliography,superscriptaddress]{revtex4-2}
\usepackage{dsfont}
\usepackage{graphicx} 
\usepackage{graphics}

\usepackage{mathtools}
\usepackage{dcolumn}    
\usepackage{bm} 
\usepackage{graphicx}
\usepackage{amsmath}    
\usepackage{latexsym}
\usepackage{amsfonts}   
\usepackage{amssymb}
\usepackage{array}      
\usepackage{epsfig}
\usepackage{txfonts}
\usepackage{xcolor}
\usepackage[colorlinks=true,linkcolor=blue,urlcolor=blue,citecolor=blue,pdfusetitle]{hyperref}

\usepackage[normalem]{ulem}

\newcommand{\norm}[1]{\left\lVert#1\right\rVert}

\newcommand{\ket}[1]{\left|#1\right\rangle}
\newcommand{\bra}[1]{\langle#1|}

\newcommand{\ketbra}[2]{|#1\rangle\langle#2|}

\usepackage[none]{hyphenat}

\newcommand{\blah}{blah\\blah\\blah\\blah\\blah}
\begin{document}
\title{Robustness of controlled Hamiltonian approaches to unitary quantum gates}
\author{Eoin Carolan}
\affiliation{School of Physics, University College Dublin, Belfield, Dublin 4, Ireland}
\affiliation{Centre for Quantum Engineering, Science, and Technology, University College Dublin, Belfield, Dublin 4, Ireland}
\author{Barış Çakmak}
\affiliation{College of Engineering and Natural Sciences, Bahçeşehir University, Beşiktaş, Istanbul, T\"urkiye}
\affiliation{TUBITAK Research Institute for Fundamental Sciences, 41470 Gebze, T\"urkiye}
\affiliation{Department of Physics, Farmingdale State College—SUNY, Farmingdale, NY 11735, USA}
\author{Steve Campbell}
\affiliation{School of Physics, University College Dublin, Belfield, Dublin 4, Ireland}
\affiliation{Centre for Quantum Engineering, Science, and Technology, University College Dublin, Belfield, Dublin 4, Ireland}
\affiliation{Trinity Quantum Alliance, Unit 16, Trinity Technology and Enterprise Centre, Pearse Street, Dublin 2, D02YN67}
\affiliation{Dahlem Center for Complex Quantum Systems, Freie Universit\"at Berlin, Arnimallee 14, 14195 Berlin, Germany}
\begin{abstract}
%~~\\
We examine the effectiveness and resilience of achieving quantum gates employing three approaches stemming from quantum control methods: counterdiabatic driving, Floquet engineering, and inverse engineering. We critically analyse their performance in terms of the gate infidelity, the associated resource overhead based on energetic cost, the susceptibility to time-keeping errors, and the degradation under environmental noise. Despite significant differences in the dynamical path taken, we find a broadly consistent behavior across the three approaches in terms of the efficacy of implementing the target gate and the resource overhead. Furthermore, we establish that the functional form of the control fields plays a crucial role in determining how faithfully a gate operation is achieved. Our results are demonstrated for single qubit gates, with particular focus on the Hadamard gate, and we discuss the extension to $N$-qubit operations.
%\\~~
\end{abstract}
\date{\today}

\maketitle

\section{Introduction}\label{intro}

Prompted by Feynman~\cite{Feynman1986}, the idea of using quantum properties of matter and light to process information has given rise to an extensive research effort. Beyond the implications for basic science, quantum information technologies would entail a significant computational speed-up for particular applications, allowing for solutions problems that are intractable with current technologies based on classical systems~\cite{Sanders2017,Preskill_2018,DeutschPRXQ}. These quantum advantages have been theoretically predicted for a variety information processing tasks, such as search and factoring algorithms, or quantum cryptography~\cite{nielsen2002quantum}. Experimentally it is now possible to implement them in systems such as superconducting qubits~\cite{Kjaergaard_2020,Rasmussen2021}, trapped ions~\cite{Bruzewicz2019}, and photons~\cite{Slussarenko2019}.

Several approaches for universal quantum computation have been developed, chief among them being measurement-based~\cite{nielsen2006cluster,Childs_2002,Childs_2005,Raussendorf_2003}, gate-based~\cite{Feynman1986,Barenco_1995}, and adiabatic models \cite{Farhi2000,Farhi_2001}. The relative benefits and drawbacks of each approach notwithstanding~\cite{Sanders2017,Preskill_2018,DeutschPRXQ}, gate-based quantum computation presents an attractive method. Any computation can be implemented by a relativity small set of gates on a qubit register~\cite{dorit2003,Barenco_1995}. Indeed, small scale quantum devices are providing remarkable platforms for simulation of quantum systems~\cite{Iskender2017,Garcia-Perez2020,Keenan2022,Guimaraes2023}, insights from which can be greatly enhanced by improving the implementation of the basic building blocks, i.e. the quantum gates.

Achieving this aim necessitates coherent control of quantum systems~\cite{Torrontegui2013,Guery-Odelin2019,Koch_2022,Carolan_2022,Sels2017,Santos_2017,IevaPRXQ}. Beyond the basic requirement of enacting the desired gate operation, we must consider several additional factors to ensure the scalability and reliability of these operations. Among these are the resources necessary for their fast and accurate implementation~\cite{tajima1,tajima2,Deffner_2021,PRXQuantum.3.020101,Yang2022,Chiribella2021,Zurek1989,Abah_2019,Cimini2020,Zhen2021,Stevens2022,Bedingham_2016,tajima3,Huber_2015}, understanding the spoiling impact of the environment~\cite{Basilewitsch2018,Koch2014,santos2021}, and the impact of operational errors~\cite{Xuereb2022,Knill_1998}. The assessment of the energetic efficiency of these devices is crucial in their design~\cite{PRXQuantum.3.020101} and may enforce practical constraints for their implementation. The interplay between the performance of a quantum computing machine and its energetic efficiency determines a fundamental connection between quantum information processing and thermodynamics~\cite{DeffnerCampbellBook,JPA_Goold}.

Following this edict, in this work we consider three approaches to implement gate operations on quantum systems through controlled Hamiltonian dynamics. In particular, we consider the auxiliary evolution approach introduced in Refs.~\cite{Hen2014,Hen2015}, where a driven {\it auxiliary} system is is coupled to the computational register upon which the operation is faithfully induced provided the evolution is adiabatic. We augment this approach with techniques from shortcuts-to-adiabaticity~\cite{Torrontegui2013,Guery-Odelin2019}, specifically counterdiabatic driving (CD)~\cite{Demirplak2003,Demirplak2005,Demirplak2008,Berry2009} and Floquet engineering (FE)~\cite{Claeys_2019}, that allow to arbitrarily speed up the implementation. In addition to these techniques, we consider an inverse engineering (IE) approach~\cite{santos2017quantum} where the computational register is {\it directly} driven by external control fields. We examine these approaches, both in terms of their resource overhead and their resilience to systematic errors stemming from imperfect timekeeping and environmental effects. We find that all are effective in achieving the target gate operation and we highlight the importance that the choice of control pulse plays in all cases. However, more significant differences emerge when considering other performance metrics and therefore we find that the optimal choice of how to realise such controlled quantum gates will ultimately be dictated by the constraints of a given architecture.

\section{Preliminaries}\label{prelim}
%%%%%%%%%%%%%%%%%%%%%%%%%%%%%%%%%%%%%%%%%%%%%%%%%%%%%%%%%%%%%
\subsection{Control protocols}
\begin{figure}[t]
\begin{center}
\includegraphics[width=0.95\linewidth]{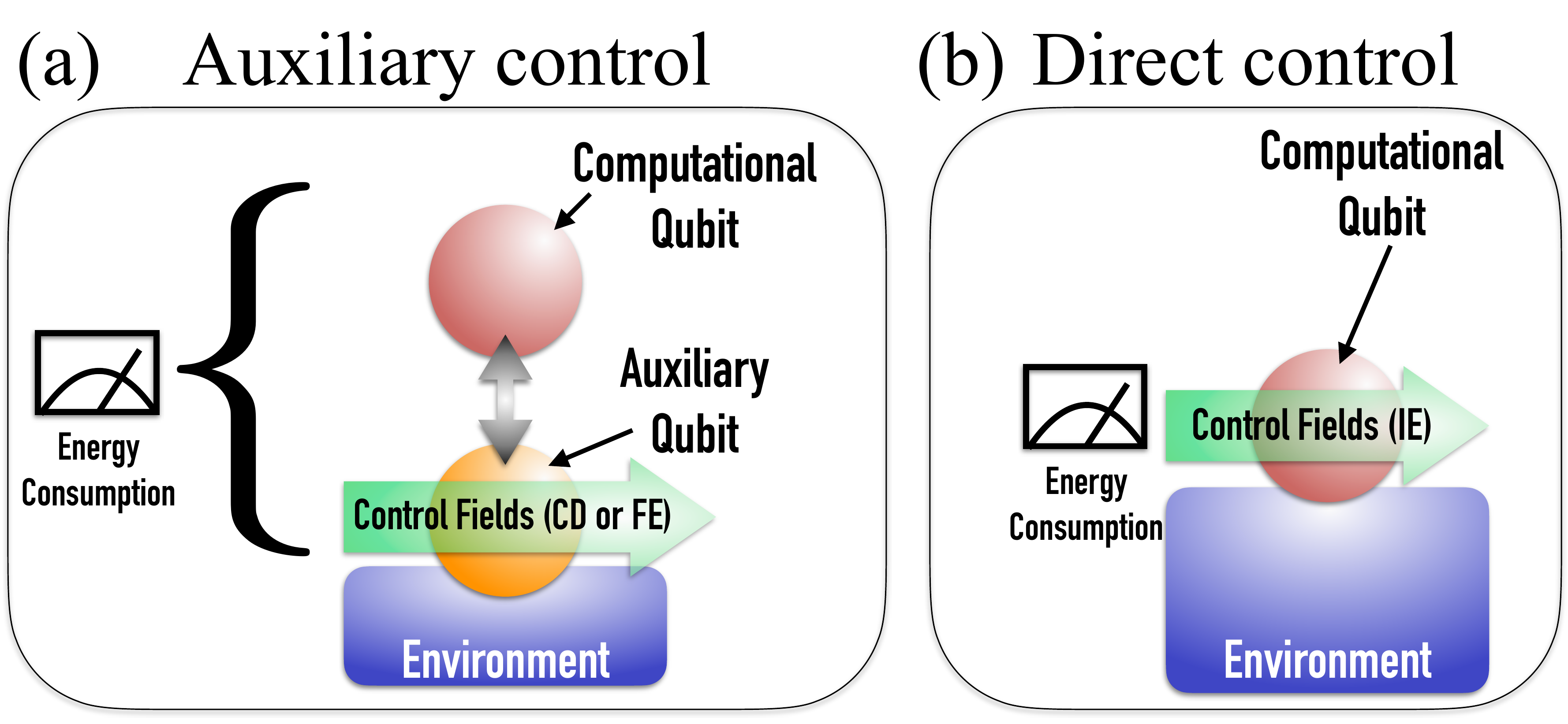}
\end{center}
\caption{(a) Shows the auxiliary control setting based on the protocol from Ref.~\cite{Hen2015}. Here an auxiliary qubit is coupled to the computational register, with control fields acting only on this auxiliary system. We assume the driven qubit can also experience environmental effects while the computational qubit is completely isolated. (b) Shows the setup for the Inverse Engineering setting. The Hamiltonian is designed without any additional resources, and thus the computational qubit is driven directly and can be subject to environmental noise.}
\label{fig_schematic}
\end{figure}

Here we outline the three control techniques that are the focus of the present work. As shown in Fig.~\ref{fig_schematic} for the auxiliary evolution approach we consider two approaches to speed up the dynamics {\it (i)} counterdiabatic driving (CD) and {\it (ii)} Floquet Engineering (FE); we also consider a third controlled implementation where the computational register is directly driven via {\it (iii)} inverse engineering (IE).

%%%%%%%%%%%%%%%SINGLE QUBIT BLOCH%%%%%%%%%%%%%%%%%%
%%%%%%%%%%%%%%%%%%%%%%%%%%%%%%%%%%%%%%%%%%%%%%%%%%
\subsubsection{Auxiliary Evolution with Control}
\label{counterdiabatic}
The first method we consider for implementing unitary gates is the adiabatic approach \cite{Hen2014} where an auxiliary qubit is coupled to a computational register upon which we wish to perform the gate operation. By driving this auxiliary qubit adiabatically, the desired gate operation can be effected on the computational register. To that end, we consider the general total Hamiltonian (for the register and auxiliary qubit)
\begin{equation}\label{genH}
    H(t)=\sum_k P_k\otimes H_{\phi_k}(t),
\end{equation}
where $P_k$ are projectors derived from the rotation axis of the desired unitary acting on the computational register, while $H_{\phi_k}(t)$ is the angle-dependent driving Hamiltonian acting on the auxiliary qubit, given by
\begin{equation}\label{gen}
    H_{\phi_k}(t)=-\left[\cos(\theta_f\lambda)\sigma_z+\sin(\theta_f\lambda)\left[\cos{(\phi_k)}\sigma_x+\sin{(\phi_k)}\sigma_y\right]\right].
\end{equation}
where $\lambda\!\equiv\!\lambda(t)$ is the time-dependent control parameter. The angles $\phi_k$ are dictated by the specific gate being implemented and we will examine an exemplary choice in the proceeding section.

This approach necessitates that the auxiliary qubit is driven adiabatically, which in general requires long timescales leaving the system open to errors from environmental effects. Counterdiabatic driving~\cite{Demirplak2003,Berry2009} allows us to arbitrarily speed up the evolution while still achieving perfect adiabatic dynamics by introducing additional term(s) to the system Hamiltonian. The CD term is equivalent to the addition of an adiabatic gauge potential to the system Hamiltonian~\cite{Sels2017}. In general, the CD term for a Hamiltonian, $H_0$, is given by
\begin{equation}
\label{cdgen}
H^{CD}=\dot{\lambda} \mathcal{A}_{\lambda}=i\sum_{m\neq n}\frac{\ket{m}\bra{m}\dot{H}_0\ket{n}\bra{n}}{E_n-E_m},
\end{equation}
where $\ket{k}$ and $E_k$ are the Hamiltonian's instantaneous eigenstates and eigenvalues. Evolving the system with the total Hamiltonian $H\!=\!H_0+H^{CD}$ gives rise to a transitionless dynamics in finite time. The exact evaluation of Eq.~\eqref{cdgen} requires the complete knowledge of the spectrum of the system Hamiltonian at all times, often limiting its applicability. However, Ref.~\cite{Sels2017,Claeys_2019} propose to approximate the exact adiabatic gauge potential given in Eq.~\eqref{cdgen} with a nested commutator expansion
\begin{equation}\label{aagp}
    \mathcal{A}_{\lambda}^{(l)}=i\sum_{k=1}^l\alpha_k\underbrace{[H[H,...[H}_{2k-1},\partial_{\lambda}H]]],
\end{equation}
where $l$ denotes the order of the expansion and, for an arbitrary system in the limit of $l\rightarrow\infty$, one obtains the exact expression given in Eq.~\eqref{cdgen}. The coefficients, $\alpha_k$, are determined by minimizing the action
\begin{align}
    S_{l}&=\text{Tr}[\hat{G}_l^2], & \hat{G}_l&=\partial_{\lambda}\hat{H}-i[\hat{H},\hat{\mathcal{A}}_{\lambda}^l].
\end{align}
This approach is particularly effective when dealing with many-body systems as it allows to truncate the complexity of the control fields~\cite{Claeys_2019}. For a single two-level system, as will be the focus of the present work, we find that Eq.~\eqref{aagp} is already identical to the full counterdiabatic term, Eq.~\eqref{cdgen}, for $l\!=\!1$, i.e. only the first term in the sum is required to achieve perfect control. However, our main interest in employing Eq.~\eqref{aagp} is because it provides a means to engineer a Floquet Hamiltonian which approximately mimics the action of adiabatic gauge potential~\cite{Claeys_2019} and therefore opens up new possibilities in terms of feasible experimental implementations~\cite{Weitenberg2021}. 

Floquet theory allows to design an effective Hamiltonian that stroboscopically mimics the dynamics of another, potentially more complex or experimentally unfeasible Hamiltonian. In order to achieve this, we need only to oscillate the original driven Hamiltonian and its derivative with respect to the driving parameter. Such a Floquet Hamiltonian can stroboscopically recreate the dynamics of the full CD Hamiltonian $H\!=\!H_0+ \dot{\lambda}\mathcal{A}_{\lambda}^{(l)}$ with a comparatively reduced operator set. The explicit form of the Hamiltonian that implements this is 
\begin{equation}\label{floqeq}
\begin{split}
    H^{FE}_{\phi}=&\left[1+\frac{\omega}{\omega_0}\cos(\omega t)\right]H_{\phi}(\lambda) \\ 
    & + \dot{\lambda}\left[\sum_{k=1}^{\infty}\beta_k\sin((2k-1)\omega t)\right]\partial_{\lambda}H_{\phi}(\lambda),
\end{split}
\end{equation}
where $\beta_k$ are Fourier coefficients of the expansion of the Floquet Hamiltonian that are chosen to match the terms of the adiabatic gauge potential expansion, $\omega_0\!=\!2\pi/\tau$ is a reference frequency and we take $\omega$ to be much greater than $\omega_0$, whose ratio defines the number of driving cycles for the evolution.

\subsubsection{Inverse Engineering}
\label{inverseeng}
As an alternative approach to auxiliary control, we consider directly driving the computational register. The evolution of a closed quantum system obeys the time-dependent Schr\"odinger equation and an arbitrary initial state is connected to a designated final state by a unitary operator, $\ket{\psi(t)}\!=\!U(t)\ket{\psi(0)}$. The Hamiltonian that generates such a unitary time evolution is determined by the well-known relation
\begin{equation}
\label{IEham}
    H(t)=i\dot{U}(t)U^{\dag}(t).
\end{equation}
It is possible to follow several approaches to inverse engineer the desired unitary~\cite{Chen2010,Kang2016,santos2017quantum,Guery-Odelin2019}, and hence the corresponding Hamiltonian. In this work, we adopt the approach taken in Ref.~\cite{santos2017quantum} and express $U(t)$ in the following form
\begin{equation}
    U(t)=\sum_n e^{i\pi\lambda_m(t)}\ket{m(t)}\bra{m(t)},
\end{equation}
where the set $\{\ket{m(t)}\}$ forms a complete orthonormal basis, and $\lambda(t)$ has the initial condition $\lambda_n(0)\!=\!2l$ where $l\in\mathbb{Z}$ to ensure $U(0)\!=\!\mathds{1}$. By taking suitable choices for the free parameters that define the orthonormal basis and local phase information, we can construct a Hamiltonian that implements the desired unitary behaviour in such a way that is not dependent on a particular initial state. In what follows we construct the IE Hamiltonian such that $\lambda(t)$ is the driving parameter. The motivation for choosing IE is to showcase another control technique. However, it is important to remark that the IE approach prescribed above and ``typical" counterdiabatic control methods are intrinsically related ~\cite{ChenPRA2011}. Thus, the results reported for the IE case would be qualitatively similar if instead CD driving were applied to the computational qubit directly. What does differ is that with the IE approach we do not start with a reference Hamiltonian {\it a priori} for which the transitions need to be suppressed. Instead, one can separate out the adiabatic gauge potential term from the resultant IE Hamiltonian after transforming basis. Nevertheless, we remind that, as depicted in Fig.~\ref{fig_schematic}, the key difference in our analysis is embodied by the two distinct settings where either an auxiliary system is employed to achieve the gate versus when the computational system is directly driven.

%%%%%%%%%%%%%%%%%%%%%%%%%%%%%%%%%%%%%%%%%%%%%%%%%%%%%%%%%%%%%
\subsection{Figures of merit}
\subsubsection{Gate Infidelity}
To characterise how faithfully a gate has been implemented we adopt the average infidelity measure~\cite{Koch2014}
\begin{equation}\label{eq:infid}
    J_T=1-\sum_{i=1}^3\frac{w_i}{tr[\rho_i^2(0)]}\mathcal{R}e\big\{tr[U\rho_i(0)U^{\dag}\rho_i(\tau)]\big\},
\end{equation}
and consider the  average of the Hilbert-Schmidt norm of the ideal evolution of three specific initial states with the obtained state, weighted by $w_i$ with $\sum_{i=1}^3w_i=1$. Three initial states satisfying particular conditions have been shown to be the minimum amount needed to address all the possible errors and characterise a general unitary operation for an open system evolution~\cite{Koch2014, Koch2013}. For a single qubit, the following set satisfies the necessary conditions~\cite{Koch2014}
\begin{align*}
    \rho_1(0)&=\begin{pmatrix}
    2/3 & 0 \\
    0 & 1/3
    \end{pmatrix},\!\!\!\!              &    
    \rho_2(0)&=\begin{pmatrix}
    1/2 & 1/2 \\
    1/2 & 1/2
    \end{pmatrix},\!\!\!\!        &
    \rho_3(0)&=\begin{pmatrix}
    1/2 & 0 \\
    0 & 1/2
    \end{pmatrix}.
\end{align*}
The first state, $\rho_1$, checks errors in the fixed basis states, and therefore does not signal any possible errors that are diagonal in this basis. The second state, $\rho_2$, addresses this and indicates the off-diagonal errors in the fixed basis. The third state, $\rho_3$, is chosen to ensure that populations are conserved, important for an open system setting. Depending on the choice of the weights in Eq.~\eqref{eq:infid}, it is possible to highlight the effect of a source of an error on the infidelity over the others. For simplicity and without loss of generality, throughout this work we choose these weights to be equal, i.e. $w_i\!=\!1/3$.

\subsubsection{Cost of Control}
The addition of control terms to the Hamiltonian implies an overall increase in resources needed to evolve the system. The analysis of this cost has been the focus of several recent works~\cite{Zheng2016,MugaNJP2018,Muga2017PRA,Abah2017,Cakmak2019,Cakmak2021,delCampoPRL2017,Chen2012,AntoPRA}, where different quantifiers have been introduced depending on the physical motivation. In this work, we adopt the cost measure introduced in~\cite{Santos2015,Zheng2016,Demirplak2008}
\begin{equation}
\label{costeq}
    \mathcal{C}=\frac{1}{\tau}\int_{0}^{\tau}\norm{H}dt,
\end{equation}
where $\norm{\cdot}$ denotes the norm of the Hamiltonian of interest, and for simplicity we consider the trace norm. It is important to emphasize that, following the approach taken in~\cite{Abah_2019}, we take $H$ to be the full Hamiltonian that generates the driven dynamics implementing the gate operation, not just the external control term. In fact, notice that it is only for the case of CD control where an explicit additional Hamiltonian term is added to the bare Hamiltonian. For both the FE and IE approaches, control is embedded into the same operators that appear in the bare Hamiltonian. Therefore, it is necessary to consider the cost of the full Hamiltonian generating the time evolution. This measure is well motivated by the functional form of the physical driving fields~\cite{Zheng2016,AntoPRA} and it has been shown to have connections to a Landauer-type limit for the change in information encoded computational states~\cite{Deffner_2021}.

\subsection{Systematic Errors}
\subsubsection{Timekeeping errors}
The controlled dynamics require that the drives are implemented for a specific length of time, which we denote by $\tau$. Since the control protocols are designed to be effective regardless of the specific functional form of the drive, this provides a useful additional degree of freedom for control protocols~\cite{DelCampo2013,Abah_2019}. We consider the following ramp profiles that satisfy the boundary conditions $\lambda(0)\!=\!0$ and $\lambda(\tau)\!=\!1$, 
\begin{eqnarray}
\begin{aligned}
\label{ramps}
&\lambda(t)=\frac{t}{\tau},& &\text{linear} \label{linear} \\
&\lambda(t)=\frac{10t^3}{\tau^3}-\frac{15t^4}{\tau^4}+\frac{6t^5}{\tau^5}, \qquad &&\text{polynomial} \label{smooth} \\
&\lambda(t)=\sin\left(\frac{\pi t}{2\tau}\right). &&\text{sinusoidal} \label{sin} 
\end{aligned}
\end{eqnarray}
We look to characterise the impact of timing errors in the drive, i.e. where the duration of the driving field over- or under-shoots the intended target time, $\tau$, by assessing the resulting impact on the gate infidelity, Eq.~\eqref{eq:infid}. We note that these pulses are chosen to capture and compare certain pulse characteristics. Indeed much work has been done in designing more complex ramp profiles via optimal control and machine learning methods~\cite{ML1,ML3,ML4} seeking to optimize to a variety of relevant cost functionals. Our analysis can therefore provide useful information for the seed pulses to ensure robustness to, e.g. time keeping errors, while exploiting more advance techniques to explore a greater optimization landscape.

\subsubsection{Environmental Errors}
We will be interested in considering how faithfully the gate operation is implemented when the controlled system is not completely isolated and therefore prone to environmental effects. To that end, we model the time evolution of the driven system with a Markovian master equation
\begin{equation}
\dot{\rho}=-i[H,\rho]+\mathcal{D}(\rho),
\label{mastereq}
\end{equation}
where $\mathcal{D}$ gives rise to a dephasing process
\begin{equation}
\begin{aligned}
 \mathcal{D}(\rho)&=\gamma(\sigma_z^{\lambda}\rho\sigma_z^{\lambda}-\rho),
\end{aligned}
\end{equation}
where the superscript $\lambda$ symbolises our assumption that the environment only affects the driven part of the system, i.e. for the CD and FE cases we assume the environment acts only on the auxiliary qubit, while for IE it is applied directly to the computational qubit(s). 

%%%%%%%%%%%%%%%%%%%%%%%%%%%%%%%%%%%%%%%%%%%%%%%%%
\section{Single Qubit Gate}
\label{singlequbit}
We begin assessing the controlled implementation of a single qubit gate which realises the operation
\begin{equation}
    \alpha\ket{n_+}+\beta\ket{n_-} \to  \alpha\ket{n_+}+\text{e}^{i\phi}\beta\ket{n_-},
\end{equation}
where $\ket{n_{\pm}}$ forms a basis in which the desired unitary gate simply performs a rotation of $\phi$. For the case of the auxiliary evolution outlined in Sec.~\ref{counterdiabatic}, where the computational qubit is coupled to an auxiliary system that is subject to the controlled drive, the combined initial state is assumed to be $\ket{\Psi_i}=\left(\alpha\ket{n_+}+\beta\ket{n_-}\right)\otimes\ket{0}$
with $\ket{0}$ and $\ket{1}$ being the eigenstates of $\sigma_z$. In this setting, Eq.~\eqref{genH} becomes 
\begin{equation}\label{eq:auxevH}
    H(t)\!=\!\ket{n_+}\bra{n_+}\otimes H_{\phi_+}(t)+\ket{n_-}\bra{n_-}\otimes H_{\phi_-}(t),
\end{equation}
where the projectors are given by $\ket{n_\pm}\bra{n_\pm}=\left(\mathbf{1}\pm\vec{n}\cdot\vec{\sigma}\right)/2$, with $\vec{n}$ being the rotation axis of the unitary and $H_{\phi_\pm}(t)$ is as given in Eq.~\eqref{gen}. For a single qubit gate, $\phi_{+}$ is taken to be zero, and $\phi_{-}$ is taken to be the angle of rotation corresponding to the desired unitary. Adiabatically evolving under the Hamiltonian above yields the following final state
\begin{equation}\label{eq:auxev1}    \ket{\Psi_f}=\alpha\ket{n_+}\otimes\ket{\epsilon_{\phi_+}^g}+\beta\ket{n_-}\otimes\ket{\epsilon_{\phi_-}^g},
\end{equation}
where $\ket{\epsilon_{\phi_{\pm}}^g}$ is the ground state of Eq.~\eqref{gen}, given by
\begin{equation}\label{eq:auxev2}
    \ket{\epsilon_{\phi_{\pm}}^g}=\cos\left(\frac{\theta_f\lambda}{2}\right)\ket{0}+\text{e}^{i\phi_{\pm}}\sin\left(\frac{\theta_f\lambda}{2}\right)\ket{1}.
\end{equation}
Choosing $\theta_f \lambda \!=\! \pi$ as the endpoint of the drive, we deterministically find that the desired gate has been implemented on the computational qubit and the auxiliary system is left in its excited state. Similarly, we explicitly show in Appendix~\ref{appdxA} that one can define a driving scheme for the auxiliary system initiated in its excited state, therefore allowing a sequence of gates to be implemented without necessitating the auxiliary qubit to be re-initialized. 

In what follows, we focus on the Hadamard gate as the target operation on the computational qubit, which corresponds to choosing the projectors as $\ketbra{n_{\pm}}{n_{\pm}}=\frac{1}{2}(\mathbf{1}\pm\frac{1}{\sqrt{2}}(\sigma_x+\sigma_z))$, accompanied by the rotations on the auxiliary qubit with phases $\phi_+\!=\!0$ and $\phi_-\!=\!\pi$, respectively, although we remark that our results are qualitatively consistent for other choices of single qubit gates. 

To drive faster than adiabatic timescales we compute the CD term, Eq.~\eqref{cdgen} for Hamiltonian~\eqref{gen} giving
\begin{equation}\label{control}
    H^{CD}_{\phi_{\pm}}(t)=\dot{\lambda}\frac{\pi}{2}[\sigma_y\cos(\phi_{\pm})-\sigma_x\sin(\phi_{\pm})].
\end{equation}
where we have taken $\theta_f\!=\!\pi$. Note that the CD term is used in addition to the bare time dependent Hamiltonian in \eqref{eq:auxevH}. It is straightforward see that the associated adiabatic gauge potential, Eq.~\eqref{aagp} is identical to Eq.~\eqref{control}, with the variational coefficient $\alpha_1\!=\!-1/4$ determined from minimizing the action, from which we readily determine the Hamiltonian giving rise to the Floquet controlled evolution
\begin{equation}\label{floqeq}
    \hat{H}^{FE}_{\phi_{\pm}}=\left[1+\frac{\omega}{\omega_0}\cos(\omega t)\right]\hat{H}_{\phi_{\pm}}(\lambda)+\dot{\lambda}\left[\omega_0\alpha_1\sin(2\omega t)\right]\partial_{\lambda}\hat{H}_{\phi_{\pm}}(\lambda),
\end{equation}
where $\omega_0\!=\!2\pi/\tau$ is the reference frequency and $\omega\!=\!N\omega_0$ with $N\!\in\!\mathbb{N}\!\gg\! 1$. The Floquet Hamiltonian replaces the time dependent bare Hamiltonians in \eqref{eq:auxevH}.

The same gate operation can be captured by the inverse engineering approach prescribed in Sec.~\ref{inverseeng}. We consider the unitary operator 
\begin{equation}
    U_1(t)=\ket{m_+(t)}\bra{m_+(t)}+e^{i\pi\lambda(t)}\ket{m_-(t)}\bra{m_-(t)},
\end{equation}
where the basis states are defined as
\begin{equation}
\begin{aligned}
    \ket{m_+(t)} &=\cos[\vartheta(t)/2]\ket{0}+e^{i\varphi(t)}\sin[\vartheta(t)/2]\ket{1}, \\
    \ket{m_-(t)} &=e^{i\varphi(t)}\cos[\vartheta(t)/2]\ket{1}-\sin[\vartheta(t)/2]\ket{0}.
\end{aligned}
\end{equation}
with parameters $\vartheta(t),\varphi(t),$ and $\lambda(t)$ that can be tuned in order to define the desired gate operation. For a single qubit gate, the driving Hamiltonian found from Eq.~\eqref{IEham} takes the form~\cite{santos2017quantum} 
\begin{equation}
    H(t)=\frac{1}{2}\vec{\omega}(t)\cdot\vec{\sigma},
\end{equation}
where the explicit form of the angular components are given in Appendix~\ref{appdxB}. The action of this Hamiltonian is to transform the input state $\ket{\mu(0)}\!=\!a\ket{0}+b\ket{1}$ to the final state $\ket{\mu(t)}\!=\!\alpha(t)\ket{0}+\beta(t)\ket{1}$ where the populations are
\begin{align*}
    \alpha(t) &=\frac{a(e^{i\pi\lambda(t)}+1)-(e^{i\pi\lambda(t)}-1)(a\cos\vartheta(t)+be^{-i\varphi(t)}\sin\vartheta(t))}{2}, \\
    \beta(t) &=\frac{b(e^{i\pi\lambda(t)}+1)+(e^{i\pi\lambda(t)}-1)(b\cos\vartheta(t)-ae^{-i\varphi(t)}\sin\vartheta(t))}{2}.
\end{align*}
One choice of parameters that gives the Hadamard gate are $\varphi(t)\!=\!0$, $\vartheta(t)\!=\!\pi/4$, and ramping from $\lambda(0)\!=\!0$ to $\lambda(\tau)\!=\!1$, which in turn gives the populations of the final state as $\alpha(\tau)\!=\!(a+b)/\sqrt{2}$, and $\beta(\tau)\!=\!(a-b)/\sqrt{2}$. The corresponding Hamiltonian that drives our qubit is then given as
\begin{equation}\label{IEHad}
    H_{Had}^{IE}(t)=\frac{\pi\dot{\lambda}(t)}{2\sqrt{2}}(\sigma_x+\sigma_z).
\end{equation}

%%%%%%%%%%%%%%%SINGLE QUBIT BLOCH%%%%%%%%%%%%%%%%%%
\begin{figure}[t]
\begin{center}
\includegraphics[width=0.85\linewidth]{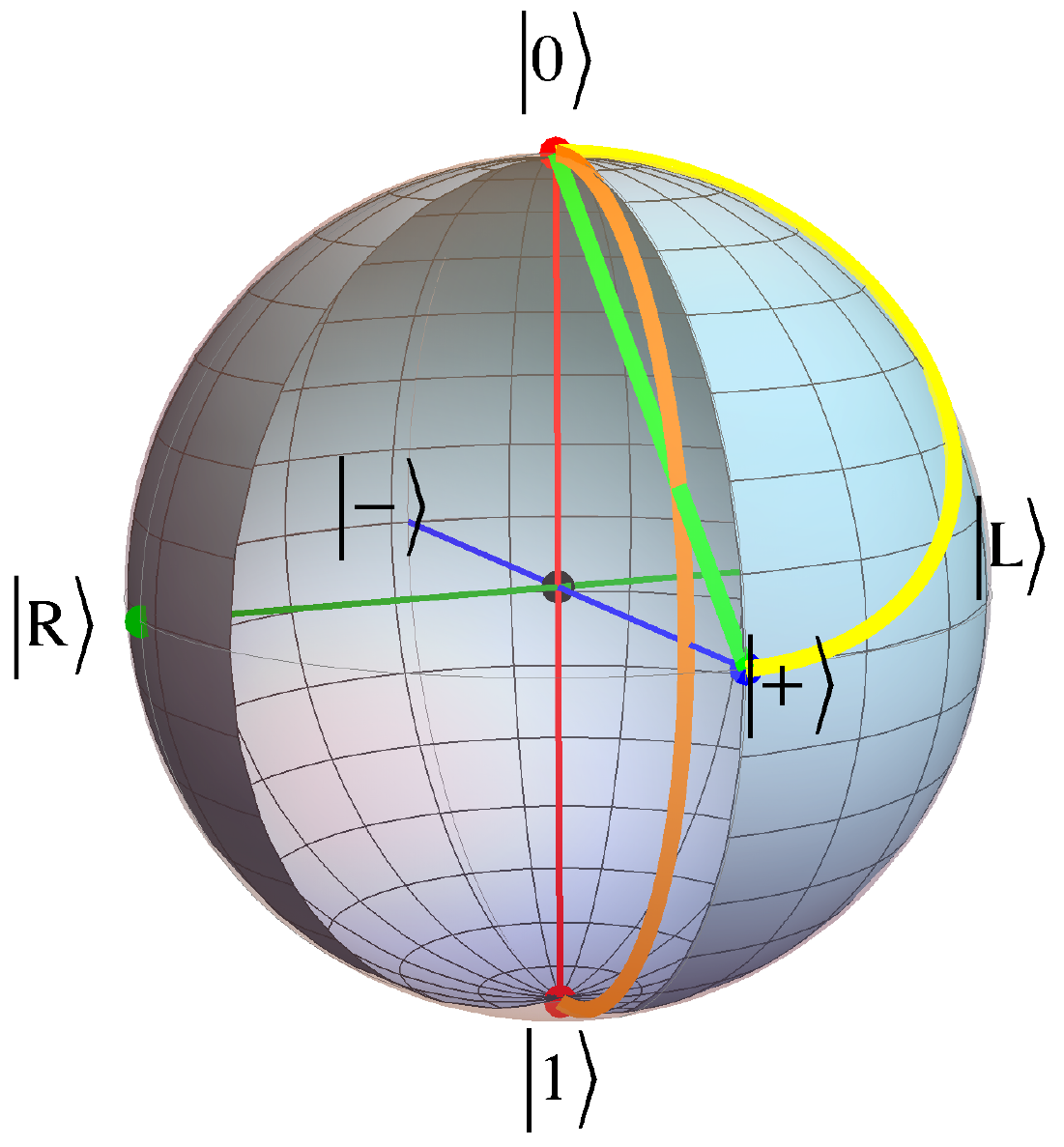}
\end{center}
\caption{We show the trajectories of the qubits for both the CD and IE protocols of the Hadamard gate. We take the initial computational state to be $\ket{+}$. The yellow line corresponds to the path of the qubit in the IE case. The green and orange lines correspond to the computational and auxiliary qubits of the auxiliary evolution cases, respectively, which begins and ends with a separable global state of the two qubits while at intermediate times the reduced states of either qubit are mixed.}
\label{fig_ideal}
\end{figure}
%%%%%%%%%%%%%%%%%%%%%%%%%%%%%%%%%%%%%%%%%%%%%%%%%%
%%%%%%%%%%%%%%%SINGLE QUBIT FIGURE%%%%%%%%%%%%%%%%%%
\begin{figure*}[t]
 {(a)} \hskip0.25\linewidth {(b)}\hskip0.25\linewidth {(c)}\hskip0.25
\linewidth {(d)}\\
\includegraphics[width=0.5\columnwidth]{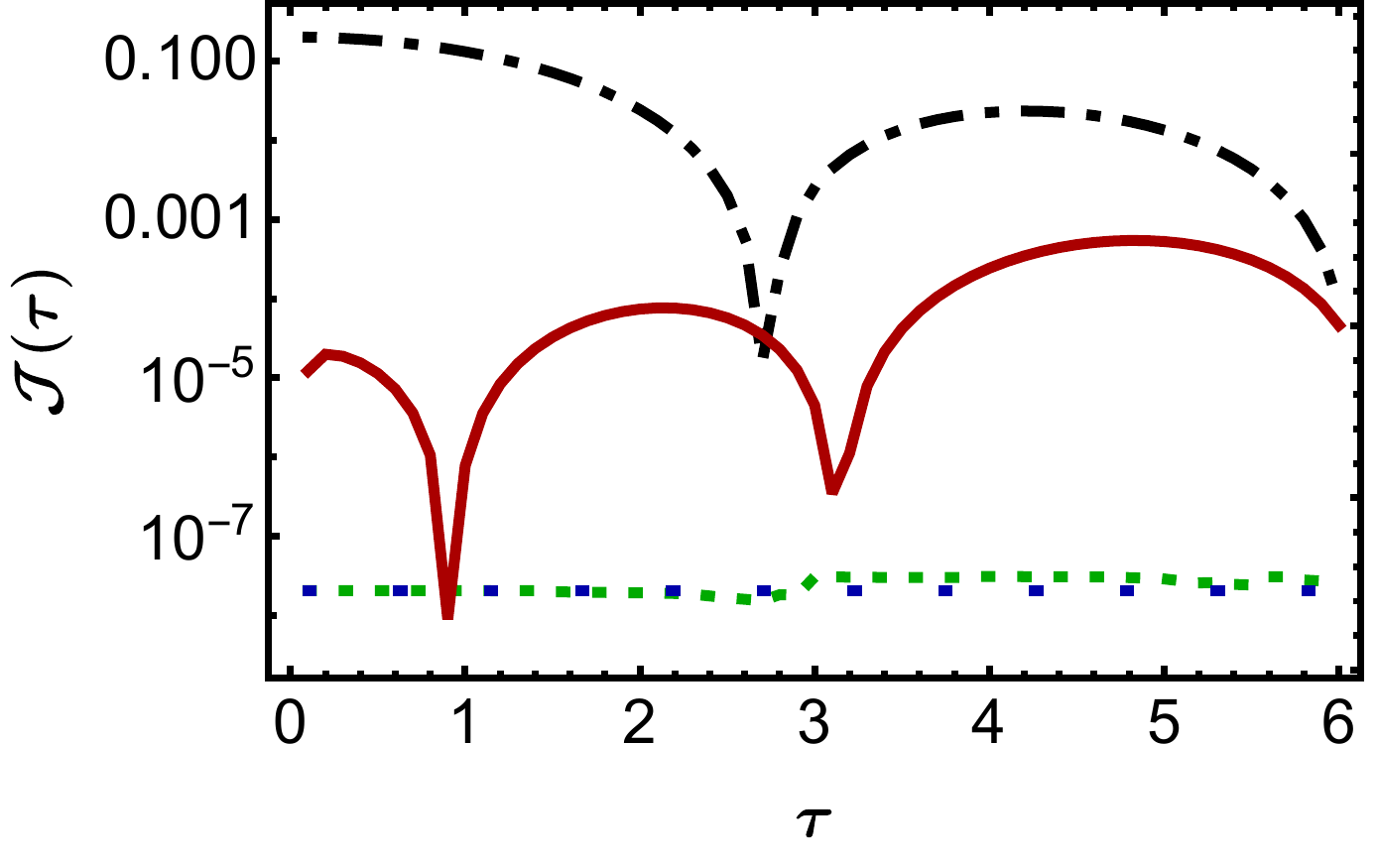}~~\includegraphics[width=0.5\columnwidth]{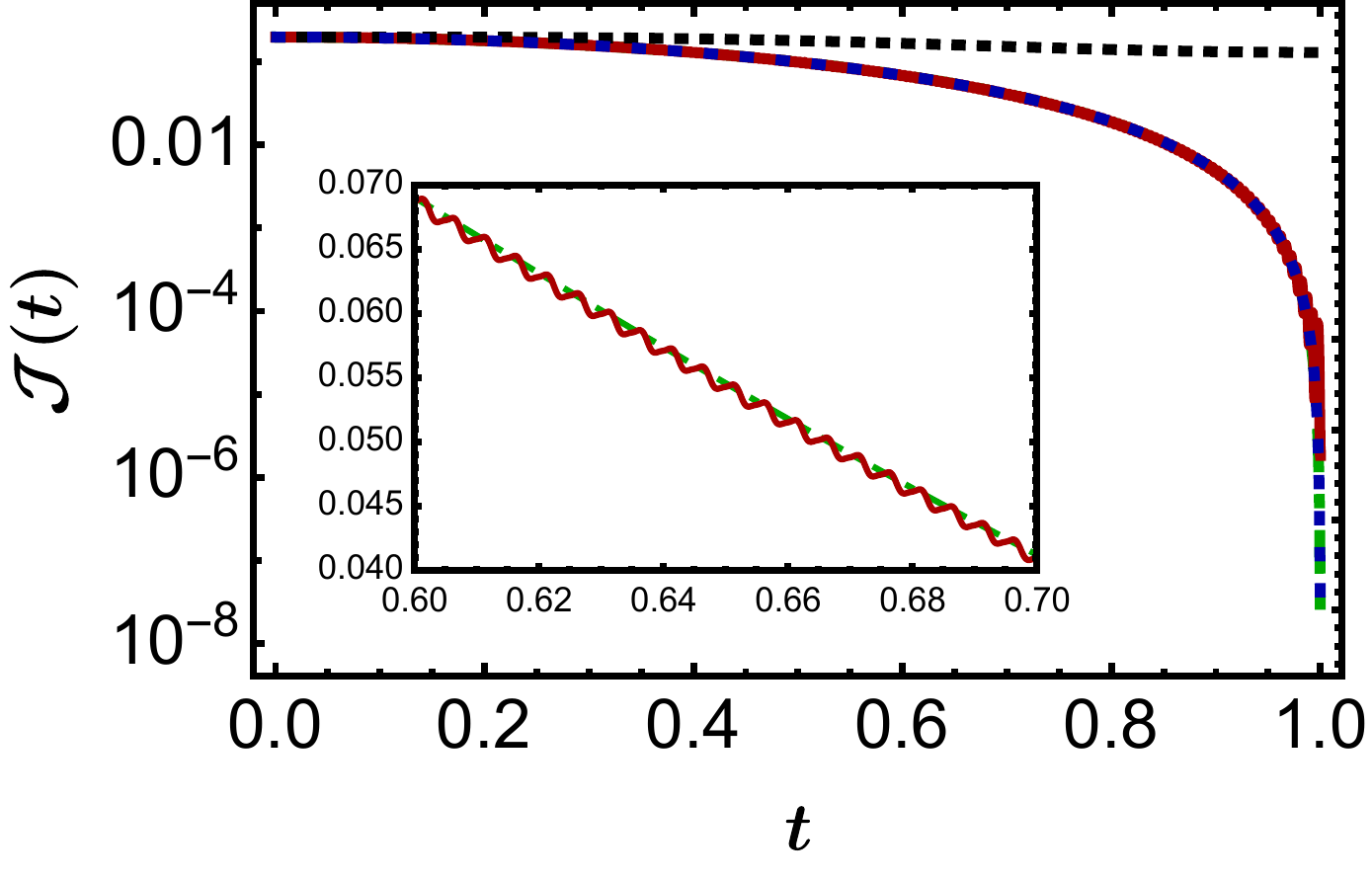}~~\includegraphics[width=0.5\columnwidth]{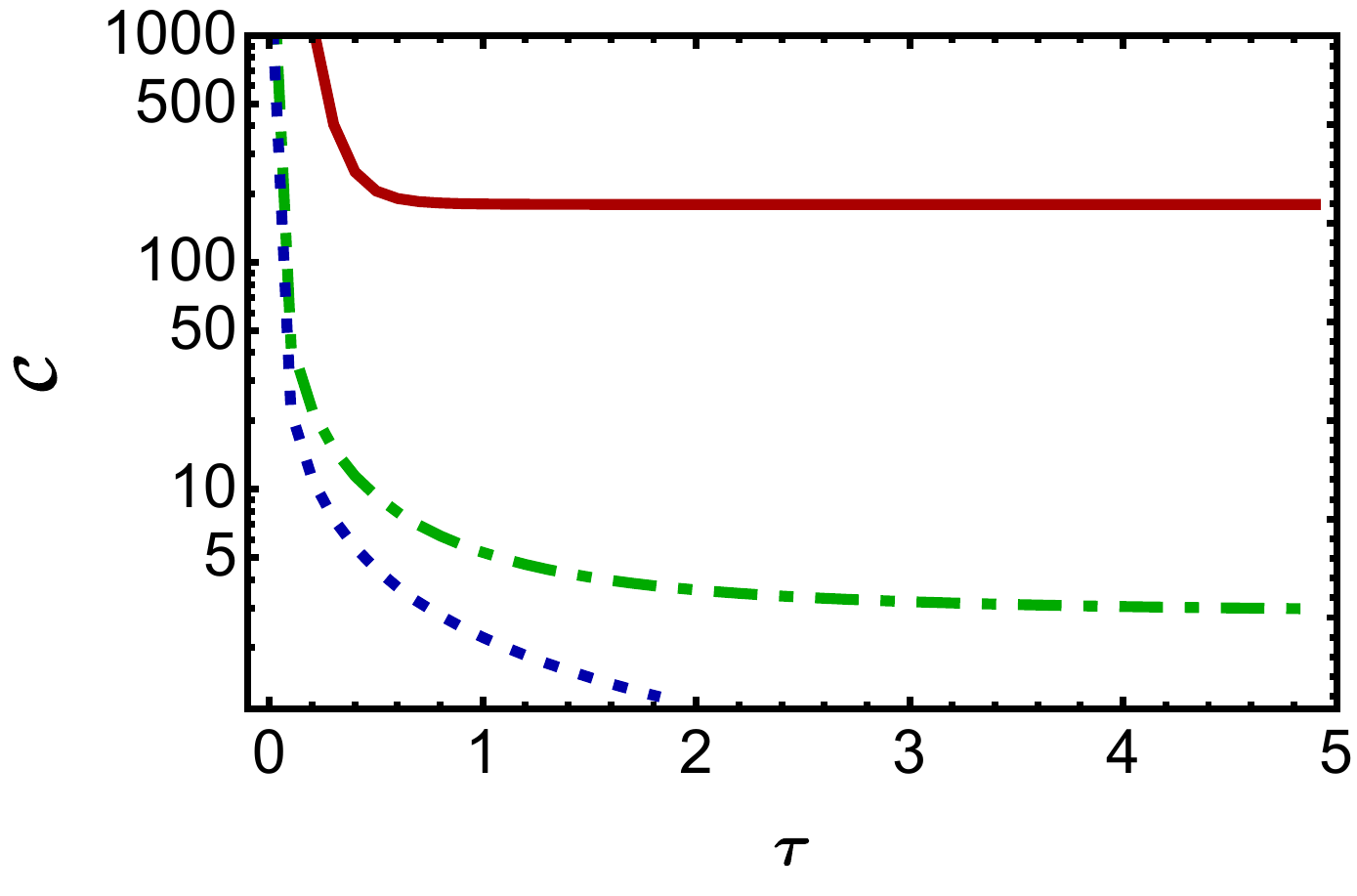}~~\includegraphics[width=0.5\columnwidth]{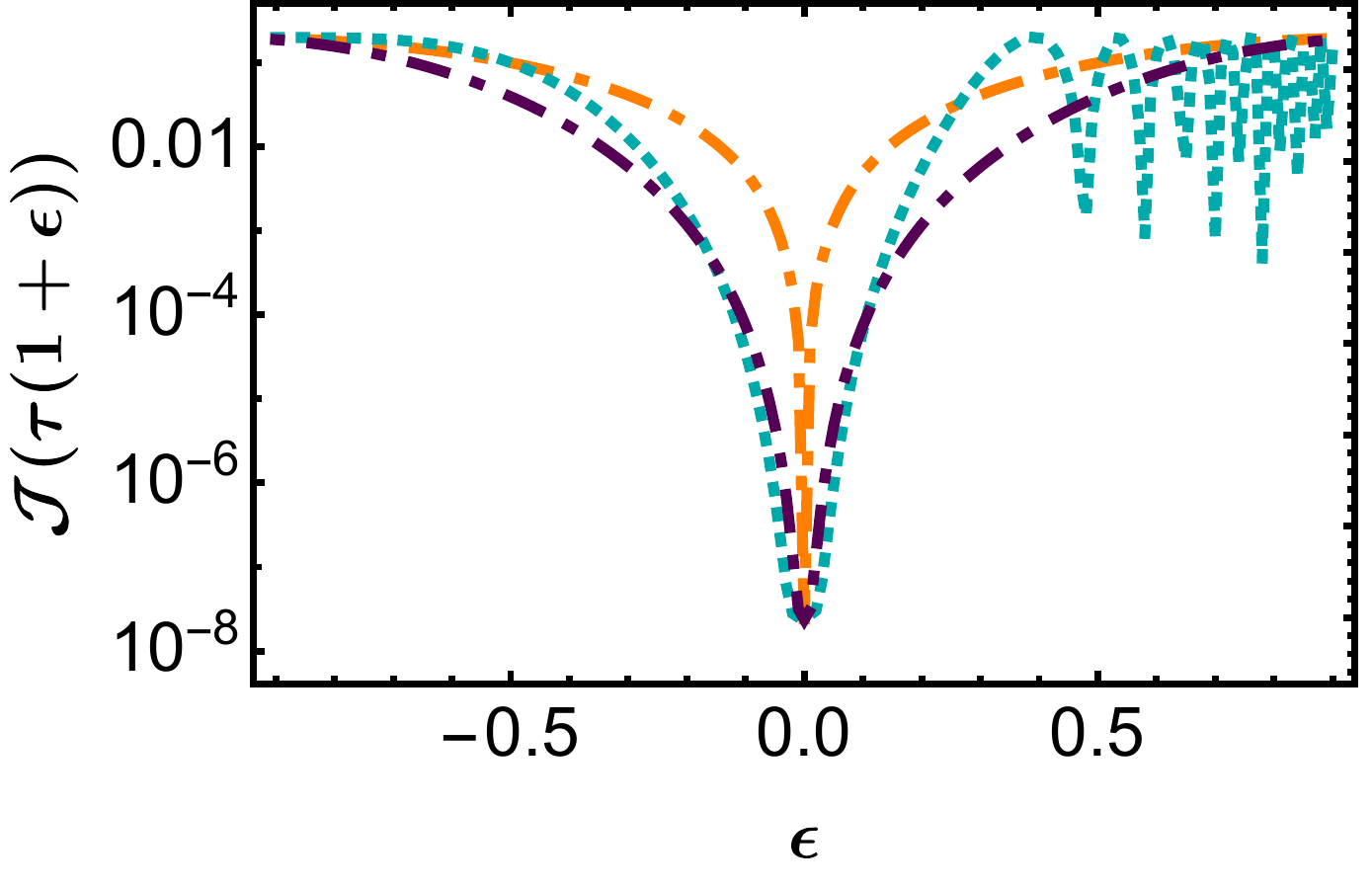}
\caption{(a) Final gate infidelity, Eq~\eqref{eq:infid} as a function of total protocol duration, for the Hadamard gate. The auxiliary evolution with counterdiabatic (CD) control is shown in the lowermost green, dotted curve. Inverse engineering (IE) performs similarly shown by the blue, dotted curve. The topmost, black, dot-dashed curve corresponds to an uncontrolled auxiliary evolution where the performance is several orders of magnitude worse. Floquet engineered (FE) auxiliary control is shown in the red, solid curve and is shown to be highly effective. (b) Dynamical gate infidelity for the Hadamard gate with $\tau\!=\!1$ using the same styling as panel (a) to identify the different control protocols. The inset captures the oscillations present in the FE driving around the dynamics of the CD approach. (c) We plot the cost, Eq.~\eqref{costeq}, of implementing the Hadamard gate vs. total protocol duration for IE, CD control, FE with $\omega/\omega_0=200$ with the same colouring as before.  (d) Final gate infidelity for the Hadamard gate vs. timekeeping error, $\epsilon$, for over- or under-shooting the intended ramp duration. The total (ideal) ramp time is $\tau\!=\!1$. We show the performance for the linear (orange), the polynomial (cyan), the sinusoidal (purple) ramps. In all panels we fix $\omega_0\!=\!2\pi /\tau$, $\omega\!=\!200 \omega_0$ for the FE case.}
\label{hadamardfig}
\end{figure*} 
%%%%%%%%%%%%%%%%%%%%%%%%%%%%%%%%%%%%%%%%%%%%%%%%%

Fig.~\ref{fig_ideal} shows the trajectories for the various control approaches on the Bloch sphere in the absence of any errors. While the IE qubit (rightmost, yellow) follows a path on the Bloch sphere and therefore remains pure during the gate operation, the auxiliary evolutions' computational qubit cuts through the Bloch sphere (straight, green line) connecting the initial state $(\ket{+})$ to the final one $(\ket{0})$. The latter observation shows that, although the initial and final states of both the control and register qubits are pure in the auxiliary evolution approach, during the dynamics they are mixed, which indicates that they become entangled during the process. By having a detailed look at the Bloch vectors of the driven qubits for the auxiliary evolution and IE, it is possible to see that their $x$ and $z$-components are equal to each other at all times, and only $y$-components differ (in fact, this component remains identically zero for the auxiliary evolution's qubit for this particular gate operation). Thus, the path that this computational qubit takes is restricted to the $x-z$ plane and the projection of the path of the IE qubit to the same plane is identical and therefore, as we demonstrate explicitly below, the performance in terms of the implementation (in)fidelity, Eq.~\eqref{eq:infid} are identical for the different processes despite their dynamics being distinct.

Fig.~\ref{hadamardfig}(a) shows the final target state infidelity for a Hadamard gate operation implemented using the three control strategies and for comparison we also show the uncontrolled auxiliary evolution (black, dashed) for a linear ramp $\lambda(t)\!=\!t/\tau$. As expected, CD and IE both achieve perfect implementations regardless of the timescale of the drive (bottom-most dotted lines). The solid red curve corresponds to the FE Hamiltonian, Eq.~\eqref{floqeq}. We see that despite the approximate nature of the FE approach, provided that the chosen parameters are within the relevant regime of applicability~\cite{Claeys_2019}, this approach is also highly effective in implementing the controlled evolution, tracking the same dynamics as the CD approach and maintaining an improvement of several orders of magnitude over the uncontrolled implementation.

In Fig.~\ref{hadamardfig}(b) we fix $\tau\!=\!1$ and examine the computational qubit's approach to the target state during the evolution. This serves to demonstrate that despite the actual dynamics giving rise to distinct paths, the effectiveness of all control protocols in terms of gate infidelity is the same. The inset demonstrates that the FE drive is a remarkably accurate approximation to the exact drive, showing small oscillations around the desired trajectory. While Fig.~\ref{hadamardfig}(a) and (b) demonstrate that, at the level of implementation, all control protocols are largely equivalent insofar as they can faithfully achieve the desired unitary, we will see in the following some qualitative differences emerge when we consider alternative performance metrics.

We show the total cost of implementing the controlled gate operation, quantified using Eq.~\eqref{costeq}, in Fig \ref{hadamardfig}(c) and for simplicity we consider a linear ramp for all protocols. To begin with, for very fast driving times, $\tau\!\to\!0$, we are in the opposite limit of adiabatic evolution and the energetic costs of all control techniques diverge. This observation is in accordance with previous works~\cite{Santos2015,Campbell2017} which establish that the energetic resources necessary to drive a system arbitrarily fast while keeping it in the adiabatic manifold requires to have access to arbitrarily large energetic resources. Naturally, for longer quench durations we asymptotically reach the adiabatic limit of the time evolution and the cost decays proportionally to $1/\tau$. Specifically, in the long time limit the CD cost asymptotically approaches to $2\sqrt{2}$, which corresponds to an unavoidable energy cost given by the energy change of bare Hamiltonian of the driven auxiliary qubit, while for IE the cost vanishes in the asymptotic limit. On the other hand, for the FE case, the leading term for the cost in the long time limit is $2\omega/\omega_0$ and proportional to the frequency of the Floquet driving, i.e. how many times the FE dynamics intersects with the true adiabatic dynamics. This requirement for high frequency driving manifests in a higher energetic cost for achieving the control.

We now turn our attention to timekeeping errors. For simplicity we focus on the case of IE, but remark the conclusions are qualitatively similar for both the auxiliary evolution cases as the dynamical overlap with the target states for the protocols coincide. In Fig.~\ref{hadamardfig}(d) we (arbitrarily) fix $\tau\!=\!1$ and consider the performance of the different ramp profiles given by Eqs~\eqref{ramps} where we allow for the ramp to over- or under-shoot the target time by a factor proportional to $1\pm\epsilon$. A simple linear ramp is the most susceptible to this type of error, with the infidelity rapidly growing as $\epsilon$ increases. Thus, while the linear ramp has some notable advantages, e.g. resulting in a time-independent control term for IE [cfr. Eq.~\eqref{IEHad}], this comes at the expense of requiring potentially costly accurate timekeeping \cite{Pearson2021}. In contrast, due to their smooth start and end points the polynomial and sinusoidal protocols allow for more severe timekeeping errors while still faithfully implementing the gate, with timing errors of up to $20\%$ still achieving infidelities $\lesssim10^{-4}$. This can be understood from the behavior of these functions at their endpoints where the rate of change of the associated driving field remains sufficiently small for $\epsilon\!<\!0.2$. As a result, the amplitude of obtaining the desired final state, which is given by $\sin^2(\theta_f\lambda/2)$ [see Eq.~\eqref{eq:auxev1} and \eqref{eq:auxev2}] does not significantly deviate from unity. These results are consistent with complementary studies of different control problems~\cite{David2022} and demonstrates that the flatness of the applied ramp around the target is an important feature to have in terms of the robustness of the protocol.

\begin{figure}[t]
\begin{center}
\includegraphics[width=0.9\linewidth]{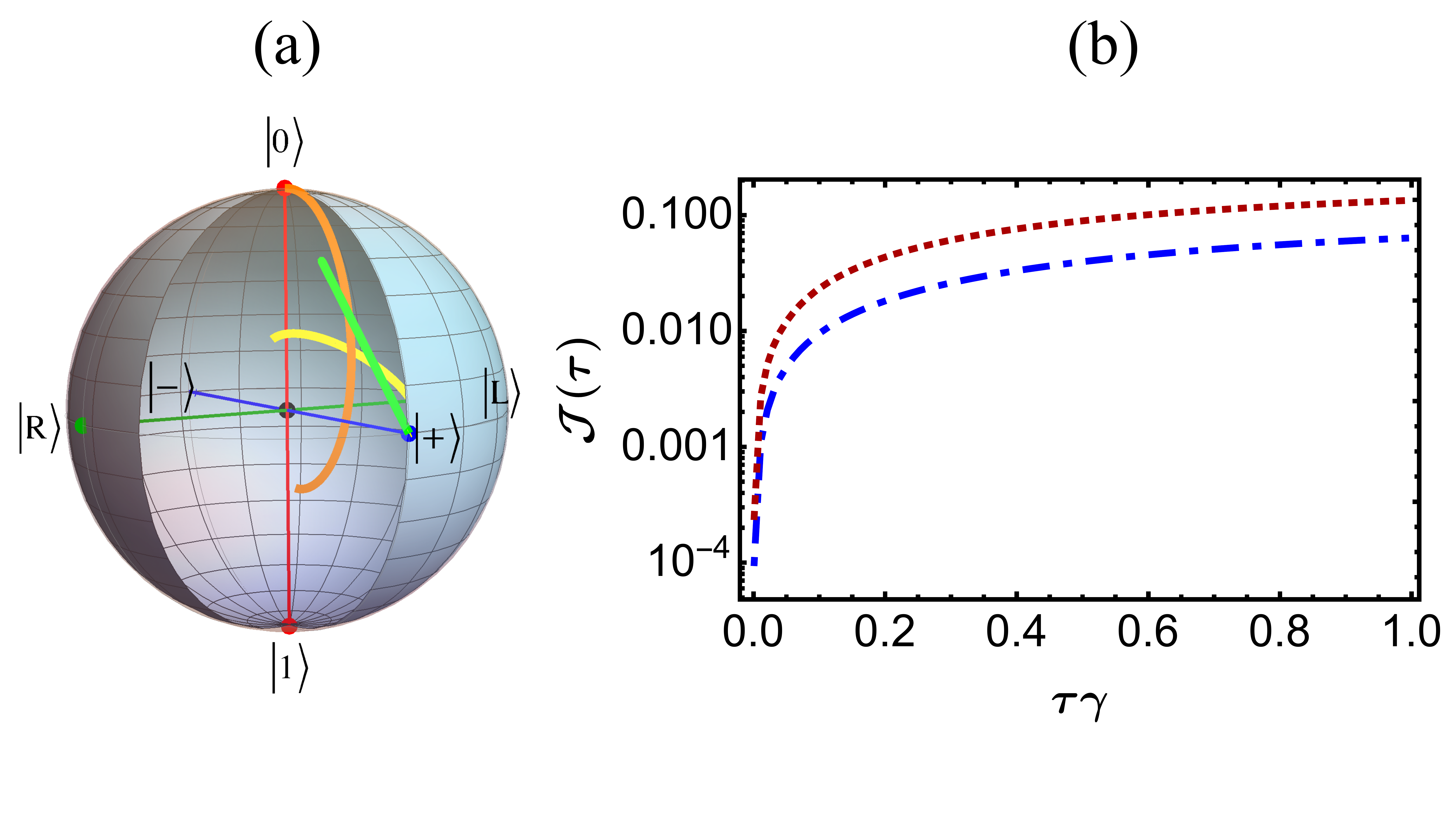}
\end{center}
\caption{(a) Qubit trajectories for the Hadamard gate under a dephasing channel, Eq.~\eqref{mastereq}, where the channel acts on the driven qubit in each case with $\tau\gamma\!=\!2$. Styling is same as in Fig.~\ref{fig_ideal}. As we dephase in the $z$-basis, both the state of the qubit in the IE case (yellow) and that of the auxiliary qubit in the CD case (orange) are pulled towards the $z$-axis. The computational qubit of the CD case (green) is not directly affected by the channel, and does not deviate from the ideal path, instead stopping along that trajectory once the auxiliary qubit driving the evolution has decohered. (b) We show the final gate infidelity, Eq.~\eqref{eq:infid}, for the dephasing channel for the Hadamard gate, with upper, dotted red and lower, dashed blue curves correspond to the IE and CD cases, respectively.}
\label{fig_density}
\end{figure}

The physical differences implied by the approaches become most apparent when considering open system effects on state evolution. Fig.~\ref{fig_density} presents our results on the infidelity between the final state and the target state as a function of the total gate implementation time, scaled with the decoherence rate for a dephasing  environment and the explicit trajectories of the qubits. In Fig.~\ref{fig_density}(a), we plot the trajectories of each qubit for the CD and IE Hadamard gate when the driven qubits are exposed to a dephasing channel. For the CD case, not driving the computational register directly can allay much of the spoiling effects of the environment. This can be understood since the auxiliary evolution approaches necessitate that the driven system ends in state $\ket{1}$, thus while the dephasing will leave the system in a mixed state, it nevertheless can have a large overlap with the intended target state of the auxiliary qubit which therefore still exhibits a good performance. Since we assume the computational qubit in the CD case does not directly feel the spoiling effects of the dephasing channel, it simply stops along its ideal trajectory when the auxiliary qubit falls short of its target state. In contrast, since we drive the computational qubit directly in the IE case while also exposing it to the dephasing channel, we see that IE qubit (yellow) starts in the $\ket{+}$ state and is drawn towards the $z$-axis, away from its ideal unitary dynamics by the environment. Fig ~\ref{fig_density}(b) shows the final state infidelity for Hadamard gate as a function of the dephasing strength. The CD case (blue) displays better final state infidelity than the IE case (red) for all values of  $\tau\gamma$. For larger gates, we expect that this difference will further widen in favour of the CD case. Despite the unfavourable cost scaling and relative complexity of the CD Hamiltonian compared to IE methods, it represents a potential attractive approach for robust gate implementation.

It is natural to consider extending the above framework to the implementation of $N$-qubit gates. The preceding analysis can readily be performed for two-qubit entangling gates, such as the controlled-phase gate~\cite{Hen2014,santos2017quantum} (we provide the explicit forms of the CD and IE Hamiltonians to implement such a gate in Appendix~\ref{appdxC}). A qualitatively similar behavior is observed: once again the overall performance in terms of process infidelity is consistent across all control approaches. Similarly, the effect of time-keeping errors is most significant for ramps that do not have smooth end points. A notable difference emerges when considering the energetic cost. While the auxiliary evolution approaches involve driving only a single qubit, and therefore the cost is essentially bounded, they can nevertheless facilitate a gate operation on an arbitrary sized register. However, this comes at the price of a difficult to realise Hamiltonian, Eq.~\eqref{genH}. This is in contrast to the IE approach where the register is controlled directly and, as might be expected, the complexity and energy required to implement IE control on multiple qubits scales poorly with the register size.

%%%%%%%%%%%%%%%SINGLE QUBIT FIGURE OPEN SYSTEM%%%%%%%%%%%%%%%%%%

\section{Conclusions}\label{conc}
We have systematically analyzed the effectiveness CD, IE, and FE methods in the Hamiltonian implementation of unitary quantum gates. As the figures of merit for all considered methods, we have put the gate infidelity, the energetic cost, susceptibility to imperfect timekeeping, and robustness against the effects of environmental noise at the center of our discussion. We have focused on the single qubit Hadamard gate and observed that all methods can faithfully achieve the desired gate, however, show some notable qualitative differences when examining performance metrics beyond target fidelities. For example, the energetic overhead of FE is the highest among the considered methods, due to the high-frequency driving necessary to achieve a gate operation closer to the ideal. As for the imperfect timekeeping errors of the desired driving time, we have observed a subtle dependence on how the Hamiltonian is driven. Smoother ramping of the Hamiltonian results in a more successful gate implementation, in case the desired driving time is over- or under-shot. Finally, we have assumed that the driven qubit in CD and IE methods is in contact with a dephasing environment, and seen that the latter control technique is more adversely affected by such environmental spoiling effects than the former due to the fact that in this case computational degrees of freedom are affected by the noise. A qualitatively similar behavior can also be observed for a finite temperature dissipative environment. We considered several commonly employed ramp profiles in order to highlight the natural robustness that each approach has under the same conditions and to provide insight into the properties that robust pulses should contain, e.g. smooth end points. This information can then be used to further enhance performance through the tailoring of ramp profiles by, e.g. optimal control techniques. However, the cost functionals can be optimized over multiple metrics, such as energetic cost, pulse bandwidth, and robustness to noise to name a few, and thus this rapidly becomes a complex problem.

We finally offer some comments on the applicability of these general Hamiltonians in light of recent experimental work has been done to implement transitionless (or superadiabatic) gates on promising candidate architectures, such as NV centres~\cite{kleissler2018universal}, superconducting qubits~\cite{wang2018experimental,wang2019experimental}, and rare-earth ions~\cite{ExpIon}. Indeed the possible universal gate sets generated by the inverse engineering case discussed in this work presents an attractive prospect for applicability, owing to the relatively simple forms and interactions present and the potential to drive them with time-independent control fields. The counterdiabatic driving case represents a departure from the typical approach to implementing a gate as it makes use of an additional auxiliary resource to mediate the driving. One may view gates in this setting as controlled gates, the Hadamard gate is perfectly implemented on the register qubit if the auxiliary qubit is driven to $\ket{1}$ and the identity is performed on it if the auxiliary is found in $\ket{0}$. This implementation therefore requires from a platform that can readily achieve controlled-gates, e.g. trapped-ion systems~\cite{Bruzewicz2019}. That this method involves inducing a phase difference between states for the computational qubit, and is more robust to noise than the direct driving approach is reminiscent of superadiabatic geometric quantum gates~\cite{liang2016proposal}. Indeed, utilising auxiliary evolution to achieve the ``superadiabatic" part of these processes could lend further robustness to these proposals.

\acknowledgements
EC acknowledges support from the Irish Research Council Project ID No. GOIPG/2020/356 and the Thomas Preston Scholarship. B\c{C} is supported by The Scientific and Technological Research Council of Turkey (TUBITAK) under Grant No. 121F246. SC is acknowledges support from the Science Foundation Ireland Starting Investigator Research Grant “SpeedDemon” No. 18/SIRG/5508 and the Alexander von Humboldt Foundation. The authors thank Anthony Kiely and Alessandro Ferraro for insightful discussions.

\appendix
\section{Excited State Driving}\label{appdxA}

For a single-qubit gate we assume that the auxiliary qubit is initially in the excited state, which implies that we initialise the system in the state
    $\ket{\Psi_i}=\left( \alpha\ket{n_+}+\beta\ket{n_-}\right)\otimes \ket{1}$.
Evolving adiabatically with the usual Hamiltonian gives the final state
\begin{equation}
    \ket{\Psi_f}=\alpha\ket{n_+}\otimes\ket{\epsilon_{0}^e}+\beta\ket{n_-}\otimes\ket{\epsilon_{\phi_-}^e},
\end{equation}
where the excited state of the time dependent Hamiltonian is
\begin{equation}
    \ket{\epsilon_{\phi}^e}=-\text{e}^{i\phi}\sin\left(\frac{\theta_f\lambda}{2}\right)\ket{0}+\cos\left(\frac{\theta_f\lambda}{2}\right)\ket{1}.
\end{equation}
Though the eigenvector has a natural $U(1)$ symmetry, a local phase appears on $\ket{0}$ due to our specification of the initial state of the dynamics at $\lambda=0$. The translation $\phi\to -\phi$ allows us to perform our gate with our auxiliary qubit initially in the excited state, returning it to its ground state at the end of the process, potentially to be reused.

\section{Explicit form of the single qubit IE Hamiltonian}\label{appdxB}

The driving Hamiltonian can be identified in the following way~\cite{santos2017quantum} 
\begin{equation}
    H(t)=\frac{1}{2}\vec{\omega}(t)\cdot\vec{\sigma},
\end{equation}
where the vector components are given as
\begin{align}
    \omega_{x}(t)=& (\cos\pi\lambda-1)\dot{\varphi}\cos\varphi\cos\vartheta\sin\vartheta \nonumber \\
    &+[\dot{\varphi}\sin\vartheta\sin\pi\lambda+(\cos\pi\lambda-1)\dot{\vartheta}]\sin\varphi \nonumber\\
    &+(\dot{\vartheta}\cos\vartheta\sin\pi\lambda+\pi\dot{\lambda}\sin\vartheta)\cos\varphi,\nonumber\\
    \omega_{y}(t)=& (\cos\pi\lambda-1)\dot{\varphi}\sin\varphi\sin\vartheta\cos\vartheta \nonumber \\
    &+[\dot{\varphi}\sin\vartheta\sin\pi\lambda-(\cos\pi\lambda-1)\dot{\vartheta}]\cos\varphi\nonumber\\
    &+(\dot{\vartheta}\cos\vartheta\sin\pi\lambda+\pi\dot{\lambda}\sin\vartheta)\sin\varphi,\nonumber\\
    \omega_z(t)&=-\dot{\vartheta}\sin\vartheta\sin\pi\lambda-(\cos\pi\lambda-1)\dot{\varphi}\sin^2\vartheta+\pi\dot{\lambda}\cos\vartheta.
\end{align}

\section{Examples of a two-qubit gate}\label{appdxC}

One typical example is the controlled-phase gate where we have two computational qubits, namely control and target, and depending on the state of the control qubit we apply a phase shift operation on the target qubit. Explicitly, if we take the initial state of the computational qubits as
\begin{equation}
\ket{\psi_i}=\alpha\ket{0,n_+}+\beta\ket{0,n_-}+\gamma\ket{1,n_+}+\delta\ket{1,n_-},
\end{equation}
then performing the controlled-phase gate yields the following output state
\begin{equation}
\ket{\psi_f}=\alpha\ket{0,n_+}+\beta\ket{0,n_-}+\gamma\ket{1,n_+}+\text{e}^{i\phi}\delta\ket{1,n_-}.
\end{equation}
we focus on the application of the controlled-$Z$ gate for which we have $\phi\!=\!\pi$. Within the auxiliary evolution framework, the gate operation can again be implemented by performing a drive on the additional ancilla qubit. This requires a Hamiltonian that implements the local phase-shift onto the $\ket{1,n_-}$ subspace while keeping the others fixed, which has the following form~\cite{Hen2014}
\begin{align}
    H_{C\pi}^{AE}(t)=&\left(\ketbra{0}{0}\otimes\mathbf{1}+\ketbra{1}{1}\otimes\ketbra{0}{0}\right)\otimes H_{0}(t) \\ \nonumber
    &+\ketbra{1}{1}\otimes\ketbra{1}{1}\otimes H_{\pi}(t),
\end{align}
where the time-dependent driving Hamiltonians $H_{0}(t)$ and $H_{\pi}(t)$ applied on the ancilla qubit are given as in Eq.~\eqref{gen}. As before, both CD and Floquet control can be used to ensure adiabatic dynamics.

Similarly we can define the same gate with IE method
\begin{equation}
    U_2(t)=\sum_{k=1,2}\ketbra{m_{k,+}(t)}{m_{k,+}(t)}+\text{e}^{i\pi\lambda_k(t)}\ketbra{m_{k,-}(t)}{m_{k,-}(t)}.
\end{equation}
The evolution basis is similarly defined
\begin{equation}
    \ket{m_{k,+}(t)}=\cos[\vartheta(t)/2]\ket{k-1,0}+\text{e}^{i\varphi_k(t)}\sin[\vartheta(t)/2]\ket{k-1,1},
\end{equation}
\begin{equation}
    \ket{m_{k,-}(t)}=\text{e}^{i\varphi_k(t)}\cos[\vartheta(t)/2]\ket{k-1,1}-\sin[\vartheta(t)/2]\ket{k-1,0}.
\end{equation}
We now have six parameters, with the restriction that $\lambda_k(0)=0$. All appear in the final state of the system under the action of the unitary. Through a suitable choice of the parameters we can design the Hamiltonian to implement the desired unitary dynamics. For example, adopting the general formalism above, we obtain the desired IE Hamiltonian that applies the controlled-$Z$ operation as follows~\cite{santos2017quantum}
\begin{equation}
    H_{C\pi}^{IE}(t)=\frac{\pi\dot{\lambda}(t)}{4}(\mathbf{1}\otimes\sigma_z+\sigma_z\otimes\mathbf{1}-\sigma_z\otimes\sigma_z).
\end{equation}
Note that the implementation of the above Hamiltonian requires a dephasing ($Z-Z$) type interaction between the target and control qubits. 

\normalem
\bibliography{bib}

\end{document}